\documentclass[letterpaper,10pt]{article}   
\usepackage{osajnl}   
 \usepackage{amssymb} 

\def\AOM {acousto-optic modulator}

\def\MOT {magneto-optical trap}

\def\ppKTP {periodically poled potassium titanyl phosphate}  

\def\SHGg  {second-harmonic generating}

\def\ARC {Australian Research Council}

\newcommand{\degC}{$^{\circ}$C}

\newcommand{\si}{$\sim$}
\newcommand{\um}{$\mu$m}
\newcommand{\uK}{$\mu$K}  
\newcommand{\uW}{$\mu$W}

\newcommand{\fastT}{$^{1}S_{0}-\,^{1}P_{1}$}    
\newcommand{\coolingT}{$^{1}S_{0}-\,^{3}P_{1}$}  

\newcommand{\Wmsq}{W\,m$^{-2}$}
\newcommand{\Yb}{$^{171}$Yb}	
\newcommand{\Ybtwo}{$^{172}$Yb}

\begin{document}

\title{Spectroscopy and laser cooling on the  \coolingT\ line in  Yb via an injection-locked diode laser at 1111.6\,nm}


\author{N. Kostylev, C. R. Locke, M. E. Tobar and J.J. McFerran$^{*}$}
\address{School of Physics, University of Western Australia, 6009 Crawley, Australia}
\address{$^*$Corresponding author: john.mcferran@uwa.edu.au}

\begin{abstract}

We generate 555.8\,nm light with sub-MHz linewidth  through the use of laser injection-locking of a semiconductor diode at 1111.6\,nm, followed by   frequency doubling in a resonant cavity.   The integrity of the injection lock is investigated by  studying an offset-beat signal between slave and master lasers, 
 by performing spectroscopy on the  $(6s)^{2}$ $^{1}S_{0}-\,(6s6p)$\,$^{3}P_{1}$ transition in magneto-optically trapped ytterbium, and by demonstrating additional laser cooling of $^{171}$Yb  with the  555.8\,nm light. For the  $^{1}S_{0}-\,^{3}P_{1}$ spectroscopy, we confirm the linear dependence between ground state linewidth and the intensity of an off-resonant  laser,  namely, that used to cool Yb atoms in a $^{1}S_{0}-\,^{1}P_{1}$ magneto-optical trap. Our results demonstrate the  suitability of injection locked 1100-1130\,nm laser diodes as a source for sub-MHz linewidth radiation in the yellow-green spectrum. 

\end{abstract}

                           


\maketitle 

\section{Introduction}

Generating narrow linewidth laser sources in the visible spectrum is a common objective, particularly in the field of atomic and molecular spectroscopy.  
   In this work we focus attention on low-noise light generation at 555.8\,nm needed for laser cooling of ytterbium, though   the scheme can be employed for wavelengths nearby; for example, those used in an iodine frequency reference~\cite{Hon2003a,Chu2005,Yam2008} or in confocal laser  scanning microscopy~\cite{Mat2011b}.   
  Laser cooling of ytterbium is becoming more prevalent as the range of atomic studies and applications increase. 
Optical lattice clocks based on ytterbium 
have achieved significant milestones~\cite{Hin2013,Lem2009a}, which has led to the rapid development of Yb lattice clocks at a number of institutions~\cite{Yas2012,Par2013,Piz2012,Nin2013}.  
Since the early magneto-optical  trapping of Yb~\cite{Lof2001}, a range of cold atoms experiments have been performed including Bose-Einstein condensation~\cite{Tak2003b,Fuk2007},  
 multi-species magneto-optical trapping with Yb~\cite{Oka2010,Mun2011,Bor2013},  photo-association spectroscopy of excited heteronuclear Yb molecules~\cite{Bor2011}, and  quantum degenerate mixtures involving Yb~\cite{Tak2009b,Har2011a,Han2013,Dor2013}.   
Investigations have been carried out studying sub-radiance in lattice bound Yb$_{2}$ molecules~\cite{Tak2012}.
Furthermore, cold ytterbium atoms or  heteronuclear Yb molecules are  good candidates for studying  topological phases
~\cite{Ger2010a,Non2013},   
searching for a permanent  electric-dipole moment~\cite{Nat2005,Tar2013}, and studying quantum magnetism~\cite{Deu2014}.  

 While methods do exist for  generating low-noise 555.8\,nm laser light, they are generally  cost disadvantageous in comparison to diode laser systems.   Furthermore,  crystal waveguides, such as those based on lithium niobate, have been known 
to suffer degradation in conversion efficiency over months or years~\cite{Chi2013a,Ste2014}. 
The technique here uses a low power fibre laser at 1111.6\,nm, then through optical injection of a semiconductor diode laser, the available power is increased while maintaining the low noise characteristic of the fibre laser.  For efficient conversion into the green a resonant frequency doubling cavity is employed. 
To confirm that the narrow linewidth of the source laser is transferred to the slave laser we perform spectroscopy on the \coolingT\ line in magneto-optically trapped ytterbium, where cooling and trapping of \Ybtwo\ is carried out using the \fastT\ transition.   We deduce that the linewidth contribution from the 555.8\,nm light is below 410\,kHz.  Furthermore, we show additional atomic cloud compression in a dual and  \MOT\  of \Yb\ involving the \fastT\ (398.9\,nm) and \coolingT\ (555.8\,nm) lines, verifying the low noise properties of the green light. 
To our knowledge, this is the first report of the use of optical injection-locking of ridge-waveguide diode lasers for narrow-linewidth green light emission near 556\,nm.  A related method has been used for 578\,nm yellow light generation~\cite{Kim2010a},  
but there the master laser  was an extended cavity diode laser locked  to a high-finesse optical cavity and a distributed feedback laser was injection locked, making for a more complicated set-up.
Single mode diode lasers in the 1100-1130\,nm range still remain scarce:  we demonstrate one of few, and perhaps the first, instance of where they are used in atomic spectroscopy.

\section{Laser injection locking at 1111.6\,nm}  

\begin{figure}[h]
 \begin{center}
{		
  \includegraphics[width=10cm,keepaspectratio=true]{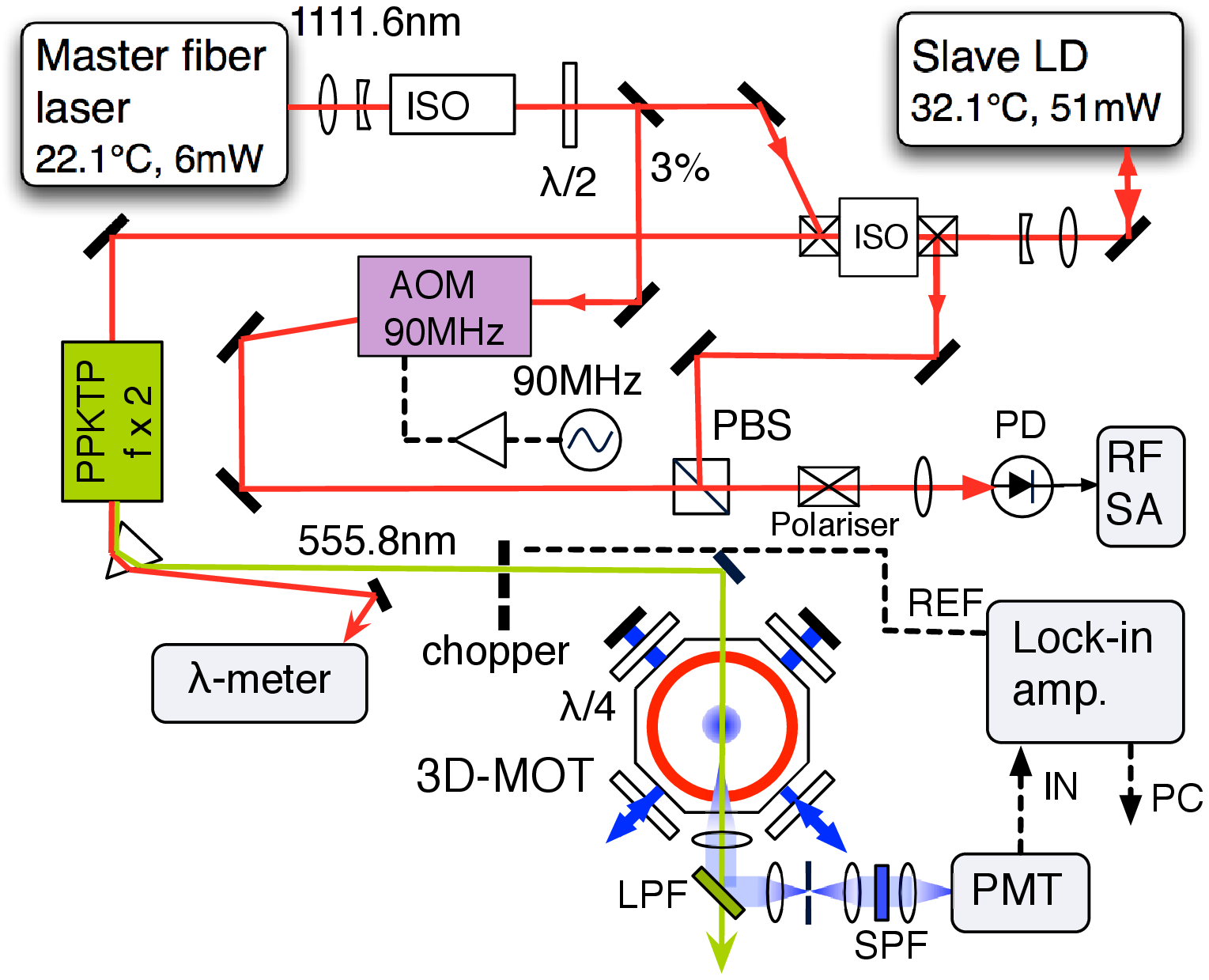}}   
\caption[]{\footnotesize     
 Experimental set up for the \coolingT\ spectroscopy of \Ybtwo\ atoms in a \fastT\ \MOT.  A  narrow linewidth (\si60\,kHz) fibre laser injection locks a semiconductor laser to increase the available power.  A few percent of the injecting light is frequency offset with an \AOM\ (f=90\,MHz) and heterodyned with the slave laser light to monitor the quality of the injection lock.  The diode laser light is frequency doubled using a \ppKTP\ (PPKTP) crystal. 
This green laser light is modulated with a rotating chopper. Fluorescence from the cold atoms de-exciting from the $^{1}P_{1}$ state is detected by a photomultiplier cell (PMT) and the modulation on the fluorescence is detected with a lock-in amplifier  referenced to the chopper frequency to recover the spectroscopic signal.  Abbreviations:   ISO, optical isolator; LD, laser diode; LPF, long pass filter at 505\,nm; MOT, \MOT; PBS, polarizing beam splitter; SA, spectrum analyser, 
SPF, short pass filter at 500\,nm).  } \label{SetUpInjection}   
\end{center}
\end{figure}

The generation of adequate power of 555.8\,nm light is often achieved by amplifying the output of a Yb doped fibre laser with a Yb fibre amplifier and then frequency doubling in a periodically poled lithium niobate crystal or waveguide~\cite{Mat2011b,Yas2010,Sin2006,Bou2005}. 
  Here we demonstrate a cost effective method that also relies on a Yb fibre laser as a master laser, but  amplification is by way of injection locking a semiconductor laser  followed by a resonant frequency doubling stage. 
 The master laser is a fibre laser  producing 6\,mW of light with a specified linewidth of less than 60\,kHz.   The slave laser is a ridge waveguide  semiconductor laser (Eagleyard, EYP-RWL-1120) with a central wavelength of 1110\,nm and a maximum power of 50\,mW.  
  The injection locking scheme is outlined 
 in Fig.\ref{SetUpInjection},  where light from the fibre laser is directed into the semiconductor laser using the rejection port of an optical isolator. 

\begin{figure}[h]
 \begin{center}
{		
  \includegraphics[width=12cm,keepaspectratio=true]{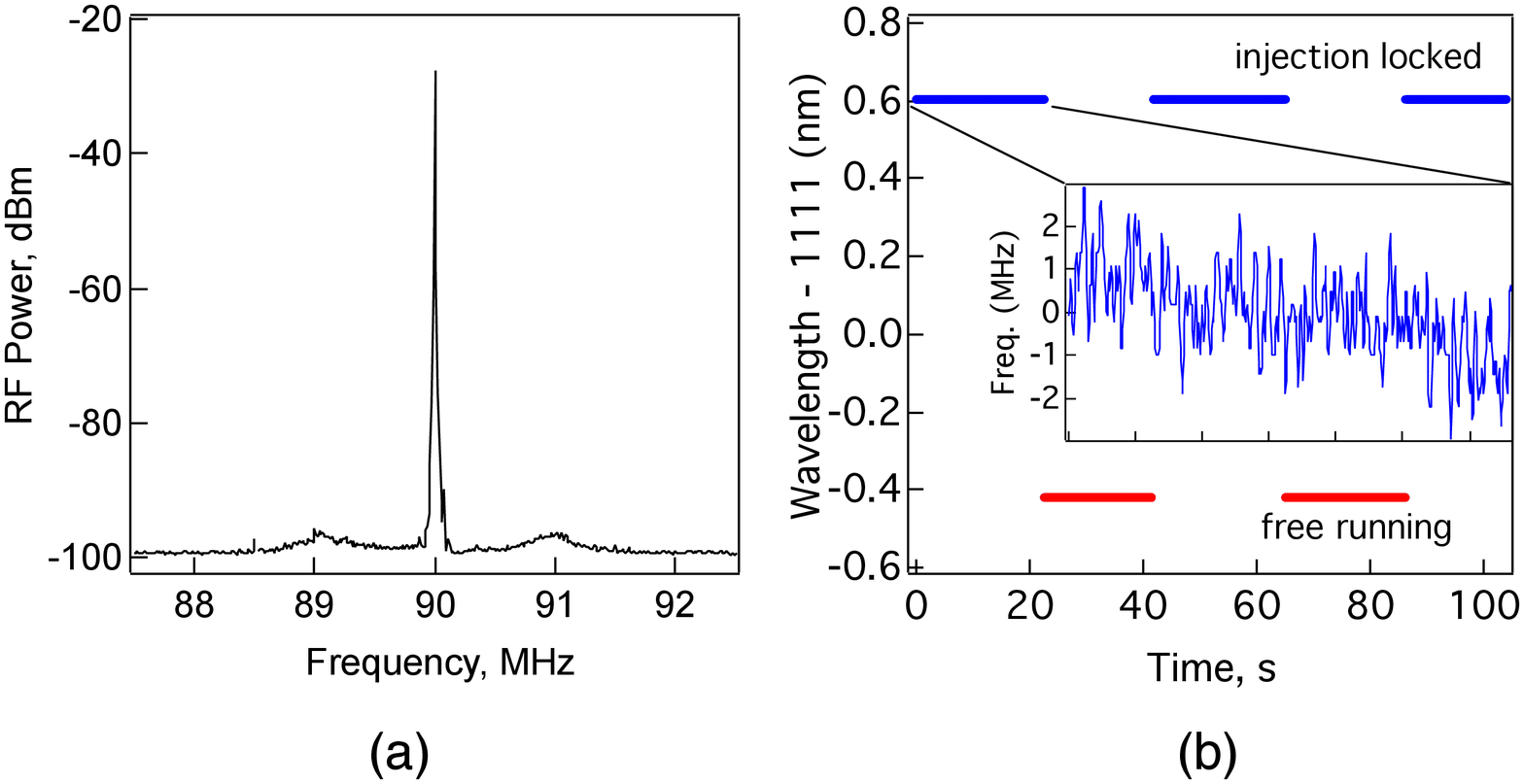}}   
\caption[]{\footnotesize     
 (a)  An optical beat signal between the slave diode laser and the  AOM-shifted master laser with 3\,kHz resolution bandwidth.  
(b)  Illustrating the wavelength shift of the diode laser by injection of light from the master laser.  The 1\,nm locking range corresponds to 240\,GHz. Inset:  The slave laser frequency, offset by the carrier frequency.
  }
   \label{InjectionLock1}  %
\end{center}
\end{figure}

  As a first means of testing the quality of the injection lock, 120\,\uW\ of the master laser light is  frequency shifted with an \AOM\ (AOM) and heterodyned with 110\,\uW\ of the slave laser light.  The resultant beat signal, at 90\,MHz, exhibits a signal-to-noise ratio (SNR) of \si70\,dB (3\,kHz resolution bandwidth) when 3\,mW of master laser light is coupled into the slave laser.   An example of the field spectrum is shown in Fig.~\ref{InjectionLock1}(a).  
   The linewidth of the optical beat signal  (\si3\,kHz) is given by that of the voltage controlled oscillator  driving the AOM. Its noise is not imposed on the semiconductor laser since the AOM lies outside the injection lock loop.  
The presence alone of the beat signal  is an indication of injection locking occurring because the diode laser is not tuneable to the wavelength of  1111.6\,nm (the nearest wavelength the diode laser could be tuned  was 1111.53\,nm after an exploration across temperature and diode current).  Furthermore, a beat signal between the lasers without injection locking produces a linewidth of approximately 20\,MHz $-$ that of the free running semiconductor laser (details below).  The residual bumps either side of the beat signal in Fig.~\ref{InjectionLock1}(a) may be due to a relaxation oscillation in the fibre laser, since the sideband frequency matches that  at which there is a 
peak in the relative intensity noise of the fibre laser.  At -70\,dB below the carrier it is not a significant concern. 

\begin{figure}[h]
 \begin{center}
{		
  \includegraphics[width=12cm,keepaspectratio=true]{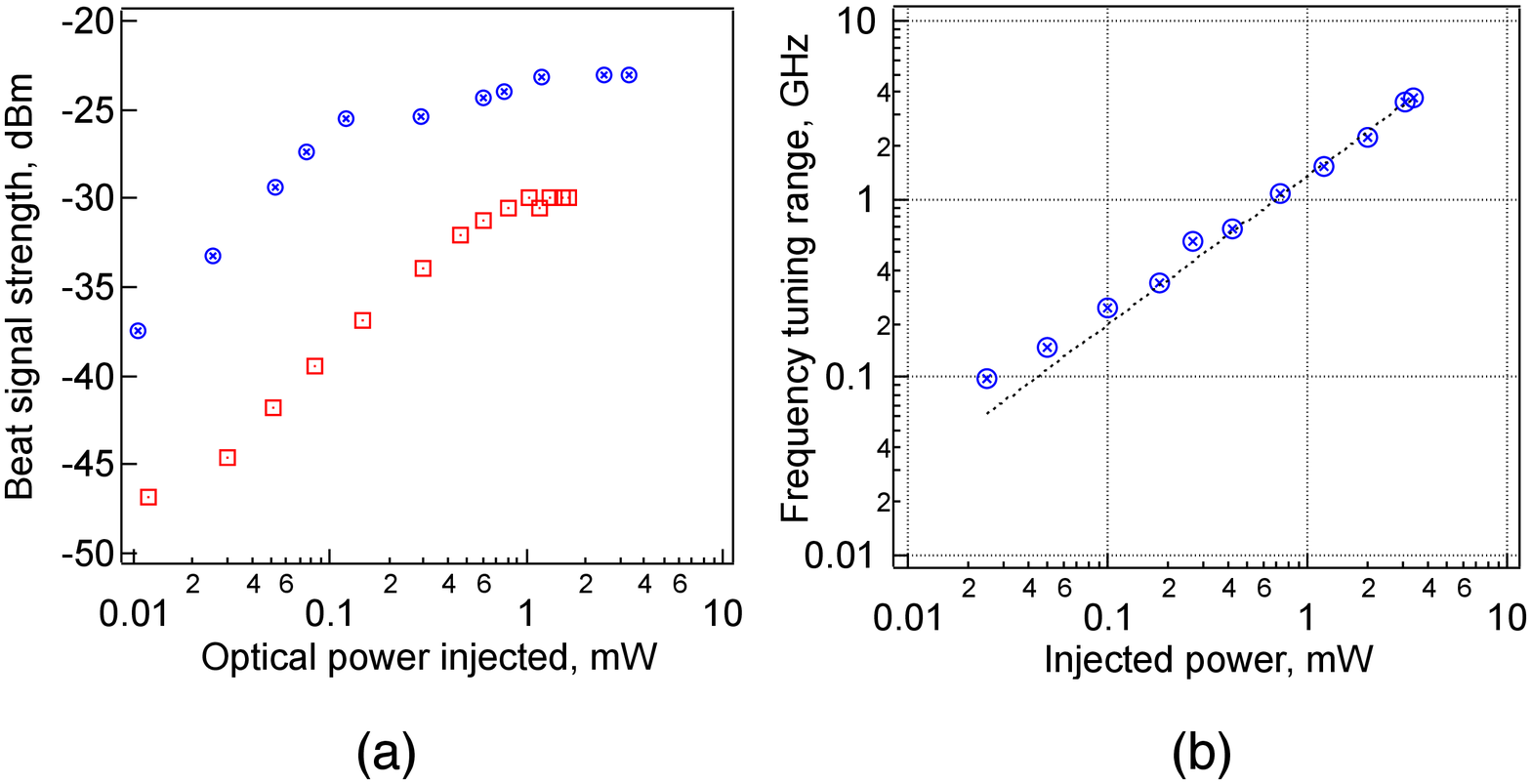}}    
\caption[]{\footnotesize    
(a) Variation of the beat signal strength versus the optical power of the injected light.
Two examples are shown where the quality of the mode-matching differs, with the upper trace having improved mode-matching,  
while the lower trace shows more distinctly the power-law behaviour. (b)  Semiconductor laser tuning range  versus the optical power of the injection locking light.  
} \label{TuningvsInjPower}
\end{center}
\end{figure}

 A demonstration of the change in wavelength of the slave laser under the influence of the master laser is seen in Fig.~\ref{InjectionLock1}(b).   Here the shift in frequency is 240\,GHz, showing the extent of the locking range.  This locking range is observed for injected light power above 150\,\uW. 
 Only at much higher resolution are the wavelength excursions seen; for example, as shown in the inset.
 The quality of the lock is discerned using the signal strength of the optical beat note. 
 In Fig.~\ref{TuningvsInjPower}(a) we show the beat signal strength versus the optical power of the injected light. 
 The beat strength increases with a power law dependence on the injected light power until a saturation level is reached, which depends on the quality of the mode-coupling into the slave laser.   In the case of optimised mode-coupling (hollow circles) the saturation power occurs at about 100\,\uW.     The lower trace (hollow squares) shows an example where the mode-coupling is not optimised, but the power law dependence is more evident.  Another power law dependence is seen between the frequency tuning range of the slave laser and the injected light power, $P$, as shown in Fig~\ref{TuningvsInjPower}(b).  
 While the locking range versus power is often used to characterise injection locking~\cite{Kob1981b},  here we consider the tuning range where the tuning is performed by adjusting a piezo element in the master laser, and the range is that where the SNR 
 of the offset-beat remains above 40\,dB (res. bandwidth =10\,kHz). There is a sharp rise in amplitude noise at the extents of the tuning range, making for a clear demarcation.  We note that the power law does not exhibit the square-root dependence sometimes seen for the locking range versus power, but rather a   \si$P^{0.8}$ dependence.  

The linewidth of the free running semiconductor laser was estimated by generating an optical beat signal between the fibre  laser and the free running diode laser, without the injection locking.  Adjustments were made to the slave laser diode temperature and current, and to the fibre laser temperature to tune the frequencies sufficiently close to one another for heterodyne detection on a photo-detector. In this case the wavelength was 1111.46\,nm.  The optical beat exhibited a linewidth of \si18\,MHz, where we can safely assume that the majority of the noise is due to the diode laser.   
Were this light to be frequency doubled  to 555.8\,nm, the corresponding linewidth  would be approximately 70\,MHz.    This may be compared with the much narrower  \coolingT\ spectra obtained below.

In demonstrating the versatility of the 1110\,nm diode lasers, 
a second semiconductor laser chip was placed in an extended cavity with a diffraction grating (1200 grooves\,mm$^{-1}$) acting as the frequency selecting element. This laser was heterodyned separately with the master fibre laser  producing an optical beat note with a linewidth of 600\,kHz (sweep time of 20\,ms).    The ratio of lengths between the extended cavity diode laser and the free running diode was \si42, while the ratio of linewidths is \si30, suggesting that the finesse is slightly higher for the diode than the extended cavity. 

%

\section{Spectroscopy of the \coolingT\ line in laser cooled Yb}
An unequivocal way of determining the resultant linewidth of the injection locked laser is to set up two independent master-plus-slave laser systems and produce a heterodyne beat.  Without access to a second fibre laser we have chosen to perform \coolingT\ spectroscopy on cold ytterbium atoms to verify the noise properties of the slave laser.
 Here we use \Ybtwo\ atoms confined in a \MOT\ (MOT) that makes use of the 398.9\,nm \fastT\ transition. The temperature of the \Ybtwo\  atoms (composite bosons) in the MOT has previously been shown to be approximately 1.1\,mK~\cite{Kos2014}.  The corresponding Doppler width, based on the expression $\Delta\nu_{D}=\nu[8\ln(2)k_{B}T/mc^{2}]^{1/2}$, is \si1.0\,MHz when probing the atoms with 555.8\,nm light.   Here, $\nu$ is the  \coolingT\  frequency, $k_{B}$ is the Boltzmann constant, $T$ is temperature and $m$ is the mass of \Ybtwo.  
The spectroscopy is performed while the MOT fields are held constant 
  and therefore the \coolingT\ is influenced by ground-state  broadening from the cooling light (sometimes referred to as Rabi broadening~\cite{Cri2010}).  
  We choose \Ybtwo\ due to the simpler energy level structure compared to the fermonic isotopes.  No dc magnetic field is applied to split the $m_{J}$ levels as the \MOT\ ensures that the $B$-field is at (or close to) zero field at the location of the atoms (we safely state this because the $m_{J=0} \rightarrow m_{J=\pm1}$ \emph{side-lobes} observed in the work of Loftus et. al.~\cite{Lof2000} are mostly unobservable, or unresolvable).  However, the linear polarisation of the 555.8\,nm light is set parallel to that of the  $B$-field  of the Zeeman slower to minimise any influence from the Zeeman slower.   Incidentally, with the 555.8\,nm light  polarised orthogonally to the Zeeman slower $B$-field  we observe splitting of the $^{3}P_{1}$ sub-states and so can verify the $B$-field strength of the Zeeman slower.    
  
  \begin{figure}[h]
 \begin{center}
{		
  \includegraphics[width=12cm,keepaspectratio=true]{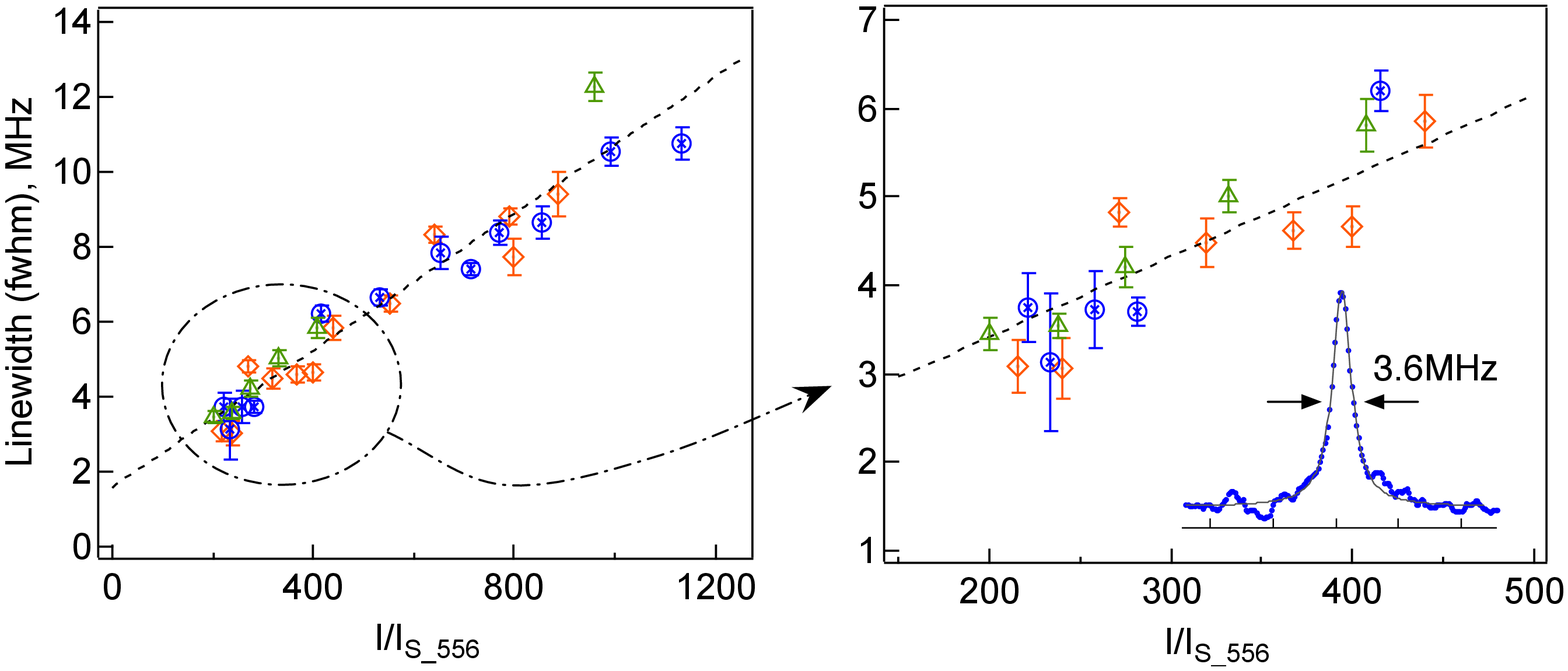}}   
\caption[]{\footnotesize     
The full-width-half-maxima of \coolingT\  \Ybtwo\ spectra versus the six-beam intensity in the \fastT\ \MOT.  The intensity is normalised using the \coolingT\ saturation intensity of 1.4\,W\,m$^{-2}$.  The different markers represent data recorded on separate days.   All data contribute to the line fit.  
Inset: \coolingT\ spectrum of the magneto-optically trapped \Ybtwo\ atoms for a 398.9\,nm cooling laser intensity of  390\,W\,m$^{-2}$ (280 $I_{S\_556}$).
  } \label{Linewidths556}    
\end{center}
\end{figure}


 The slave  laser light is frequency doubled using a \ppKTP\ (PPKTP) crystal producing \si3.5\,\uW\ of 555.8\,nm light with 36\,mW of fundamental light incident. 
  The poling period is 10.275\um\ (Raicol Crystals specification) and the crystal temperature for optimum frequency doubling is 55.7\degC\  (the FWHM temperature bandwidth is 3.5\degC).  
 The 555.8\,nm probe beam, with 2.4\,\uW\ of optical power sent to the MOT chamber, is modulated with a chopper wheel rotating at 379\,Hz.  The probe creates a depletion in the ground state and a corresponding modulation is produced in the 398.9\,nm fluorescence of the cold atoms.  The fluorescence is detected with a photomultiplier cell (Hamamatsu 10492-001, 1\,V/$\mu$A, bandwidth of 20\,kHz) and the modulation is down-converted with a lock-in amplifier that is referenced to the chopper wheel frequency.  A short-wavelength pass filter with an optical density of 6.0 at 556\,nm prevents any stray green light reaching the PMT.   The 398.9\,nm light was intensity stabilised via an amplitude modulation servo that uses the same AOM that controls the violet light's frequency detuning.  The technique differs from previous \coolingT\ spectroscopy in a ytterbium MOT, where  556\,nm fluorescence was detected and the measurements were recorded at DC~\cite{Lof2000,Cri2010}. 
An example of the spectra obtained here is shown in the inset of Fig.~\ref{Linewidths556}, in this case, with a MOT light field intensity of 390\,W\,m$^{-2}$.
 The frequency scale is determined with the aid of a wavemeter that continually records the wavelength of the 1111.6\,nm light (and the frequency doubling is taken into account).  A piezo element in the fibre laser permits scanning of the laser frequency. 
 The time taken to sweep across the FWHM of the resonance is \si4\,s and the maximum drift rate of the master laser is 2.1\,kHz\,s$^{-1}$ (4.2\,kHz\,s$^{-1}$ at 556\,nm), therefore the drift  does not strongly influence the spectral width of the lines. Any influence is reduced by scanning in alternate directions across the transition and calculating a mean FWHM. 
The dependence of the spectral-width of the \coolingT\ line as a function of the optical intensity  in the MOT is shown in Fig.~\ref{Linewidths556}.  The intensity is normalised by the saturation intensity  of the \coolingT\ transition; i.e., $I_{S\_556}=1.4$\,\Wmsq.  
 The different markers represent data recorded on separate days.  Each data point  is the weighted mean of between three and five measurements, where each measurement is the mean FWHM of typically four  line scans. 
 The Zeeman slower beam intensity was reduced to \si120\,$I_{S\_556}$ for these measurements; however,  the  \coolingT\ linewidth was found to be independent of the Zeeman beam intensity, so is not included in the intensity values of Fig.~\ref{Linewidths556}.
 There is a clear reduction in linewidth as the intensity is reduced.  
The relationship  often used to describe intensity broadening is
 $\Delta\omega_{m}(I)  = \Gamma(1+I/I_{S})^{1/2}$, where $\Gamma$ is the natural linewidth, or more generally, the homogenous linewidth in the absence of light~\cite{Sle2010,Aku1990}. 
 However, here the experiment is arranged such that only the ground state of the \coolingT\ transition is affected by strong intensity perturbations.
The 556\,nm light intensity was $\approx$ 0.012 $I_{S\_556}$, so its contribution to line-broadening is negligible.  
  Line broadening of the ground state is characterised by a linear dependence on intensity~\cite{Coh1998,Cri2010}. 
 Along with the natural linewidth there are also other contributions to the zero-intensity linewidth; for example, the  inhomogeneous broadening from the residual velocity of the cold atoms ($\Delta\nu_{D}$), and noise from the 555.8\,nm light ($\Delta\nu_{L}$).   Incorporating these into the expression for spectral-width we have:    $\Delta\nu_{m}(I) \approx \Delta\nu_{D} + \Delta\nu_{L}+ (\Gamma/2\pi) (1+ bI/I_{S\_556})$, where $b$ characterises the slope. 
   The dashed line of Fig.~\ref{Linewidths556} shows the weighted line fit, producing an ordinate intercept of $1.59\pm0.09$\,MHz (all the data are used to produce the fit).  The product-moment correlation coefficient (Pearson's $r$ coefficient) is 0.89, supporting the linear  relationship between the linewidth and the intensity.   Subtracting linewidth contributions due to Doppler broadening ($\Delta\nu_{D}=1$\,MHz) and the natural linewidth ($\Gamma_{556}=182$\,kHz), leaves a contribution from the 555.8\,nm light of $\Delta\nu_{L}\approx410$\,kHz.  
The 1111.2\,nm fibre laser has a linewidth specification of 60\,kHz; therefore   the linewidth at 555.8\,nm should be about 240\,kHz (making the tenable assumption that the laser noise is dominated by white frequency noise).   Our measurement is reasonably  consistent with this estimate.  
 In fact, it should be treated as an upper bound on the laser's linewidth since residual $m_{J=0} \rightarrow m_{J=\pm1}$ splitting may to contribute to the linewidth if the  atom cloud  is offset from the MOT centre (and this is more likely to occur when the MOT light fields are weak).
We note that the temperature of the Yb atoms increases with the cooling laser intensity; therefore, the Doppler broadening contribution to the linewidth will also increase.  However, here the corresponding maximum linewidth (at \si1200 $I_{S\_556}$) is only \si1.6\,MHz based on Doppler cooling theory (at a frequency detuning equal to 1.5 times  the linewidth of the \fastT\  transition).    Thus, the contribution to the linewidths in Fig.~\ref{Linewidths556} from this effect is small (and, moreover, has a  $I^{1/2}$ dependence). 
Although we did not detect it, one may expect some ground state-broadening from the Zeeman slower beam, which would also lead to an overestimate of the laser linewidth (the slowing beam frequency detuning is -7\,$\Gamma_{399}$ from the centre of  the \fastT\ transition, but this should not prevent it from perturbing the $^{1}S_{0}$ state). 
Reduced uncertainty on the zero-intensity spectral-width would be achieved if the 398.9\,nm laser cooling light could be reduced further. However,
below 200 $I_{S\_556}$ (\si300\,W\,m$^{-2}$ or 0.5 $I_{S\_399}$) there is insufficient cooling intensity to trap adequate numbers of atoms in the MOT for \coolingT\ line detection.

Another test regarding the injection locking is to reduce the power of the light injected into the slave laser diode and observe if there is an influence on the width of the   \coolingT\ spectral line when probing the cold atoms. 
We found that there was no noticeable change in the spectral width versus injected power and that only the strength of the spectroscopic signal weakens as the injected power is reduced.  The control bandwidth of the injection locking is therefore well beyond the range of several MHz as one might expect~\cite{Tel1993}.
   The change in strength of the \coolingT\  signal with injected power implies that   the fraction of slave laser light that is coherent with the master laser also changes with injected power.  Below a threshold of about 1\,mW,   
the fraction of slave laser light that is coherent with the master laser light reduces with a logarithmic dependence: in our case with a
 slope of $\approx0.05$\,dBm$^{-1}$. 

%

\section{Magneto-optical trapping with 555.8\,nm light}

To enhance the level of 555.8\,nm light the PPKTP crystal has been placed in a resonant frequency doubling cavity~\cite{Uet2008}.  The cavity is designed to generate a 35\,\um\ waist at the crystal location in both vertical and horizontal planes.   The radius of curvature of the folding mirrors is 75\,mm and full length of the cavity is 568\,mm.  
The cavity resonances further demonstrate the effectiveness of the injection locking, since clear Lorenztian lineshapes are resolved by the cavity.  Without injection locking the resonances are heavily obscured.  
The cavity is readily locked to the centre of  resonance with the frequency modulation method ($f_{\mathrm{mod}}$=33\,kHz).  We confer a 4.5\,kHz servo bandwidth between the \SHGg\ cavity and the 1111.8\,nm light (which may be restricted by the phase roll-off produced by the lock-in amplifier used to generate the error signal).

\begin{figure}[h]
 \begin{center}
{		
  \includegraphics[width=9cm,keepaspectratio=true]{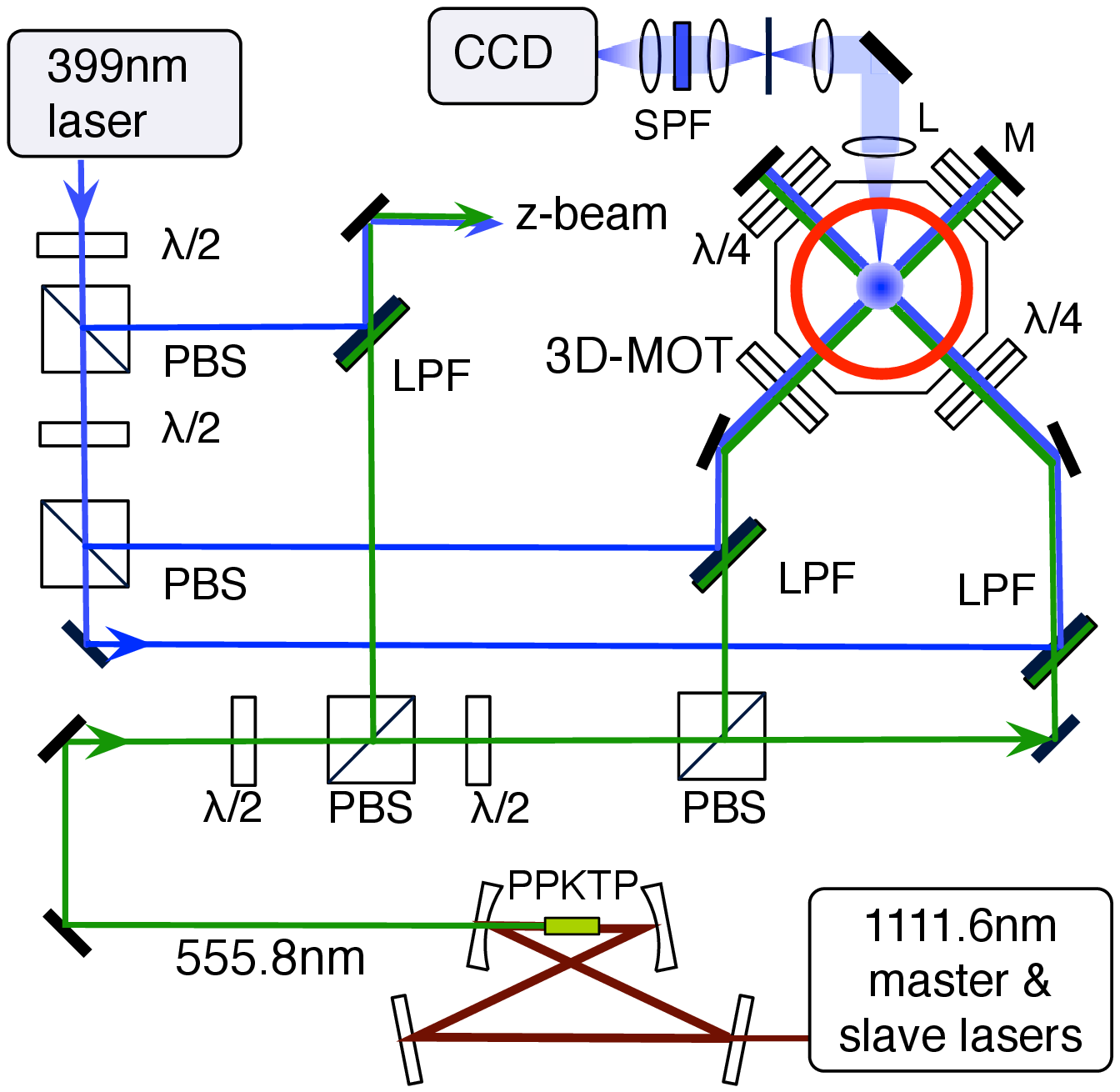}}    
\caption[]{\footnotesize    
 Sketch of the set up for the dual \fastT\ and \coolingT\ \MOT.  The 398.9\,nm and 555.8\,nm beams proceed along the same paths through the MOT. Abbreviations: CCD, charged coupled device; L, lens, LPF, long pass filter at 505\,nm; M, mirror, MOT, \MOT; PBS, polarizing beam splitter; PPKTP, \ppKTP; $\lambda/4$, dual wavelength  (399\,nm \&\ 556\,nm) quarter wave plate; SPF, short pass filter at 505\,nm; z-beam, refers to the beam along the \emph{strong}-axis of the MOT.
  } \label{DualMOT}  
\end{center}
\end{figure}

The   beam profile  of the  diode laser light is set with two pairs of cylindrical lenses for improved coupling into a single mode optical fibre. 
The fibre spatially filters the light to aid with the mode-matching into the frequency doubling cavity, and transports the light to the magneto-optical trap table.  There is 52\,\% of the light coupled through the fibre, making available 24\,mW for the resonant cavity.   The  spherical-Gaussian beam profile out of the fibre helps to achieve  98.5\,\%  mode matching into the TEM00 mode of the \SHGg\ cavity. 
By observing the reflection from the input coupler, the fraction of the 1112\,nm light coupled into the cavity  is, at most, 58\,\%.
We see improved coupling as the incident power increases.  It follows that the impedance matching is improving as the nonlinear conversion increases.
The optimum reflectivity of the input coupler can be approximated using $R_{\mathrm{opt}}  \approx 1-P_{c}\gamma-\epsilon$, where $P_{c}$ is the circulating power, $\gamma$ is the single pass nonlinear conversion efficiency (0.026\,W$^{-1}$)~\cite{Koz1988}, and $\epsilon$ represents other losses of fundamental power in the cavity excluding that at the input coupler~\cite{Let2005,Koz1988}.  It is evident in our case that $\epsilon$ is very low because the finesse of  the cavity corresponds closely to the losses produced by the transmission of the input coupler ($R=0.93$).   Making up the difference we find  $\epsilon\sim0.01$. 
  The power enhancement is about  28; hence, for an input power of 25\,mW the optimum input coupler reflectivity is  \si0.96 (the power enhancement varies by \si14\,\% over $0.93<R<0.99$).  
Should diode lasers at 1110\,nm become available with higher power the impedance matching condition will be better met.   Alternatively, if an input coupler with R=0.96 is used an increase of about  30\,\%  in the level of 555.8\,nm light may be obtained. 
We can readily produce 8\,mW of  555.8\,nm light $-$ when half of this light is incident upon the atoms in the \fastT\ MOT, the atoms are completely ejected from the trap.
Moreover, for a beam with an $e^{-2}$ radius of 5\,mm, a power level of 8\,mW corresponds to an intensity 150 times  the saturation intensity of the \coolingT\ transition.

\begin{figure}[h]
 \begin{center}
{		
  \includegraphics[width=12cm,keepaspectratio=true]{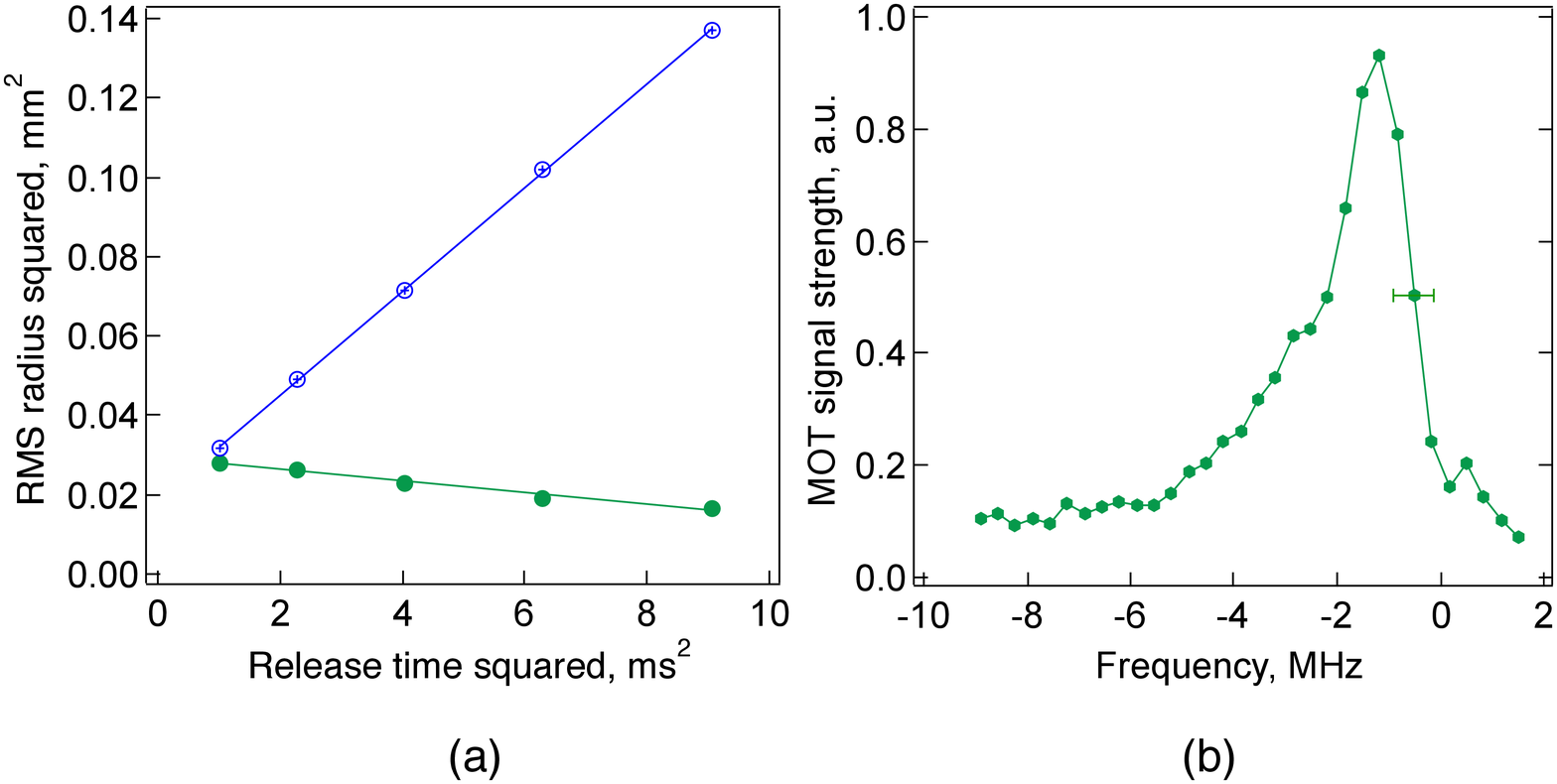}}    %
\caption[]{\footnotesize    
 (a) The root-mean squared cloud radius versus free expansion time (the 398.9\,nm  light and MOT $B$-field gradient are turned off).   The upper trace is without the 555.8\,nm light present;  the lower trace is with the 555.8\,nm light present and red frequency detuned with respect to the \coolingT\ transition.  The slope of the upper trace corresponds to a temperature of 270\,\uK. (b) Cold atom fluorescence as a function of 555.8\,nm frequency detuning, after a release time of 2.0\,ms from the 398.9\,nm MOT.   
  } \label{GreenMOT}  
\end{center}
\end{figure}

To confirm the narrow linewidth of the 555.8\,nm  light, we have tested its suitability  for laser cooling of Yb, where  we have implemented a dual stage \MOT\ using dichroic filters to combine the 398.9\,nm and 555.8\,nm beams, and achromatic quarter wave plates for the polarisation control. The arrangement is outlined in Fig.~\ref{DualMOT}.  The overlapping beam paths at the two wavelengths produces a compact configuration and eases the alignment process for the 555.8\,nm light.  The photo-multiplier tube of Fig.~\ref{SetUpInjection} is now replaced with a CCD camera to image the cloud of atoms. In Fig.~\ref{GreenMOT}(a) we show two traces: in the upper trace the root-mean-squared (rms) cloud size is measured against the free expansion time of the \Yb\ atoms (both  the 398.9\,nm light and magnetic gradient are turned off for the free expansion).  In the lower trace the same occurs, but with the 555.8\,nm beams passing through the MOT and red-detuned from the centre of the \coolingT\ resonance (by approximately one linewidth).  We see  that the green light further compresses the cloud of atoms, indicating further cooling.  This supports earlier results that the injection locking process is preserving the narrow linewidth of the fibre laser.  

As a final demonstration we show, in Fig.~\ref{GreenMOT}(b),  the enhancement of the cold atoms'  fluorescence when the 555.8\,nm is tuned close to the \coolingT\ transition. While the relative frequency scale is well calibrated, the absolute frequency with respect to the line-centre is not, but it is  coarsely estimated from the known characteristics of laser cooling.  The fluorescence is recorded 2\,ms  after both the 398.9\,nm light and magnetic gradient are turned off. The cycle time and therefore time taken between data points is 1\,s, and the plot was produced by averaging over six scans.  The asymmetric line shape is characteristic of spectra obtained with \MOT s $-$ the sharp roll-off near resonance reflecting the sudden transition from trapping to dispersing atoms.  The linewidth of approximately 1.6\,MHz is enhanced by the broadening produced by an intensity of \si190\,$I_{\mathrm{S}\_556}$.  The effectiveness of the 555.8\,nm radiation is evident.

\section{Conclusions}

We have demonstrated a scheme of generating low-noise 555.8\,nm light suitable for \coolingT\ Yb laser cooling that uses injection locking of a semiconductor laser at 1111.6\,nm. A low power fibre laser injection locks a 50\,mW  diode laser, after which second harmonic generation  in a resonant cavity is carried out. It avoids the use of high-power fibre amplifiers and crystal waveguides,  
both of which are costly and the latter
can suffer degradation in conversion efficiency over the long term~\cite{Chi2013a}.   The technique should make experiments with laser cooled Yb more accessible.  We performed spectroscopy on the \coolingT\ line in \Ybtwo\ with 0.01 $I_{\mathrm{S}\_556}$ of 555.8\,nm light and showed a linear dependence between the ground state linewidth and intensity of the \fastT\ trapping light.   Interpolating to zero intensity we infer that the linewidth of the 555.8\,nm light is less than 410\,kHz.  As further confirmation of the 555.8\,nm light properties, we have shown additional laser cooling of \Yb\ atoms in a dual 398.9\,nm$-$555.8\,nm \MOT, hence indicating the suitability of injection locked 1100-1130\,nm laser diodes as a source for sub-MHz linewidth radiation in the yellow-green spectrum.

\section*{Acknowledgement}
This work was supported by the \ARC\ (LE110100054).  J.M. is supported through an ARC Future Fellowship (FT110100392) and N.K. through a Prescott Postgraduate Scholarship, UWA.   We are gracious to Gary Light and Steve Osborne of the UWA Physics workshop for their technical expertise.   We thank members of the ARC Centre of Excellence for Engineered Quantum Systems  for their assistance, and S. Parker and E. Ivanov for the use of equipment.




%


\end{document}